\def\useextern{}

\documentclass[sigconf]{acmart}

\usepackage[nameinlink,noabbrev,capitalize]{cleveref}
\usepackage{caption}
\usepackage{subcaption}

\ifluatex
\usepackage[newfloat,finalizecache]{minted}
\else
\usepackage[newfloat,frozencache]{minted}
\fi
\newminted{c}{}
\newminted{cpp}{}
\newminted{text}{}
\newmintinline{c}{style=bw}
\newmintinline{cpp}{style=bw}
\newmintinline{text}{style=bw}

\usepackage{pgfmath}

\ifdefined\useextern

\usepackage{tikzexternal}
\else
\ifdefined\genextern

\usepackage{tikz}
\usetikzlibrary{external}
\else

\usepackage{tikz}

\fi

\usetikzlibrary{calc}
\usetikzlibrary{decorations}
\usetikzlibrary{decorations.pathreplacing}
\usetikzlibrary{decorations.text}
\usetikzlibrary{decorations.pathmorphing}
\usetikzlibrary{decorations.markings}
\usetikzlibrary{fit}
\usetikzlibrary{shadows.blur}
\usetikzlibrary{arrows}    
\usetikzlibrary{shapes}
\usetikzlibrary{shadows}
\usetikzlibrary{automata}
\usetikzlibrary{positioning}
\usetikzlibrary{petri}
\usetikzlibrary{chains}
\usetikzlibrary{fadings}
\usetikzlibrary{matrix}
\usetikzlibrary{patterns}
\usetikzlibrary{mindmap}
\usetikzlibrary{graphs}			
\usetikzlibrary{graphdrawing}	
\usetikzlibrary{quotes}			
\usetikzlibrary{babel}          
\usetikzlibrary{hobby}
\usetikzlibrary{arrows.meta}
\usetikzlibrary{backgrounds}
\usetikzlibrary{bending}

\usegdlibrary{trees}			
\usegdlibrary{layered}			
\usegdlibrary{force}			
\usegdlibrary{circular}			
\usegdlibrary{routing}			

\pgfdeclarelayer{backbackbackground}
\pgfdeclarelayer{backbackground}
\pgfdeclarelayer{background}
\pgfdeclarelayer{foreground}
\pgfdeclarelayer{foreforeground}
\pgfsetlayers{backbackbackground,backbackground,background,main,foreground,foreforeground}
\tikzset{>={Stealth[round,flex,length=5pt 4.5 0.8]}} 
\tikzset{tight/.style={minimum width=0pt,minimum height=0pt,inner sep=0pt,outer sep=0pt}}

\fi

\input{macros}

\acmConference[ICPP '21]{ICPP '21: 50th International Conference on Parallel Processing}{August 09--12, 2021}{Chicago, IL}
\acmBooktitle{ICPP '21: 50th International Conference on Parallel Processing, August 09--12, 2021, Chicago, IL}

\begin{document}

\title{Loop Transformations using Clang's Abstract Syntax Tree}

\author{Michael Kruse}
\orcid{0000-0001-7756-7126}
\affiliation{%
  \institution{Argonne National Laboratory}
  \streetaddress{9700 S. Cass Avenue}
  \city{Lemont}
  \state{Illinois}
  \country{USA}
  \postcode{60439}
}
\email{michael.kruse@anl.gov}

\begin{abstract}
OpenMP 5{.}1 introduced the first loop nest transformation directives \texttt{unroll} and \texttt{tile}, and more are expected to be included in OpenMP 6{.}0.
We discuss the two Abstract Syntax Tree (AST) representations used by Clang's implementation that is currently under development.
The first representation is designed for compatibility with the existing implementation and stores the transformed loop nest in a shadow AST next to the syntactical AST.
The second representation introduces a new meta AST-node \texttt{OMPCanonicalLoop} that guarantees that the semantic requirements of an OpenMP loop are met, and a \texttt{CanonicalLoopInfo} type that the OpenMPIRBuilder uses to represent literal and transformed loops.
This second approach provides a better abstraction of loop semantics, removes the need for shadow AST nodes that are only relevant for code generation, allows sharing the implementation with other front-ends such as flang, but depends on the OpenMPIRBuilder which is currently under development.
\end{abstract}

\ccsdesc[500]{Software and its engineering~Compilers}
\ccsdesc[300]{Software and its engineering~Parsers}
\ccsdesc[500]{Software and its engineering~Parallel programming languages}
\ccsdesc[300]{Software and its engineering~Software performance}

\keywords{OpenMP, Clang, abstract syntax tree,
semantic analysis,
code generation}

\maketitle


\section{Introduction}

\begin{figure}
\begin{resizepar}
\large\sffamily

\begin{tikzpicture}
\tikzset{layerhighlight/.style={line width=4pt,draw=green!80!black,top color=green}}
\tikzset{filemanagerlayer/.style={}}
\tikzset{sourcemanagerlayer/.style={}}
\tikzset{lexerlayer/.style={}}
\tikzset{preprocessorlayer/.style={}}
\tikzset{parserlayer/.style={}}
\tikzset{semalayer/.style={}}
\tikzset{codegenlayer/.style={}}
\tikzset{astlayer/.style={}}
\tikzset{>={Straight Barb[angle=60:12pt,width=12pt,round] }}

\tikzset{bubble/.style={draw=gray,fill=white,inner sep=1pt}}
\tikzset{node/.style={draw=teal,line width=1.2pt,rounded corners,top color=teal!50,bottom color=white,shading angle=15,blur shadow}}
\tikzset{layer/.style={draw=teal,line width=1.2pt,rounded corners,top color=teal!50,bottom color=white,shading angle=320}}
\tikzset{layerlabel/.style={xslant=1,font=\bfseries,anchor=base west}}
\tikzset{journey/.style={draw=teal!80!black,line width=4pt,->}}
\tikzset{steering/.style={draw=yellow!80!black,line width=2pt,->}}
\tikzset{hole/.style={circle,fill=black,xslant=0.4,inner sep=2.5pt}}
\tikzset{ast/.style={circle,bubble,astlayer}}
\tikzset{data/.style={ellipse,bubble}}
\tikzset{file/.style={shape=file,fill=white,draw=black}}

\node[file] (ir) at (4,-2) {source.ll};
\draw[journey] (4,0) -- (ir);

\path[layer, codegenlayer] (0,0) -- +(2,1.8) -- +(9.8,1.8) -- +(8,0) -- cycle; 
\path[layer] (3.5,-0.5) -- +(1.2,01.2) -- +(4.2,1.2) -- +(3,0) -- cycle; 
\node[layerlabel,font=\small\bfseries] at (4.4,0.3) {OpenMPIRBuilder};
\path[layer] (2.5,-0.5) -- +(0.6,0.6) -- +(3.4,0.6) -- +(2.8,0) -- cycle; 
\node[layerlabel,font=\small\bfseries] at (3.3,-0.3) {IRBuilder};
\node[hole] at (4,0.4) {};
\draw[journey] (4,1.6) |- (-0.6,1.4);
\draw[journey] (-0.6,1) -| (4,0.4);

\path[layer,semalayer] (0,1.6) -- +(2,1.8) -- +(9.8,1.8) -- +(8,0) -- cycle; 
\node[hole] at (4,2) {};
\draw[journey] (4,3.2) -- (4,2);
\node[ast] at (5.2,2.4) {AST};
\draw[journey] (5.6,2.8) -| (6.8,2.4) node[right,font=\small,rotate=-90,anchor=south,align=center]{Tree-\\Transform} |- (5.6,2);

\path[layer,parserlayer] (0,3.2) -- +(2,1.8) -- +(9.8,1.8) -- +(8,0) -- cycle; 
\node[hole] at (4,3.6) {};
\draw[journey] (4,3.6) -- (4,5);
\draw[journey] (0.6,5.2) -| (-0.8,6.8) |- (0.2,8.2);

\path[layer,preprocessorlayer] (0,4.8) -- +(2,1.8) -- +(9.8,1.8) -- +(8,0) -- cycle; 
\node[hole] at (4,5.2) {};
\draw[journey] (4,5.2) -- (4,6.6);

\path[layer,lexerlayer] (0,6.4) -- +(2,1.8) -- +(9.8,1.8) -- +(8,0) -- cycle; 
\node[hole] at (4,6.8) {};
\draw[journey] (4,6.8) -- (4,8.2);

\path[layer,sourcemanagerlayer] (0,8) -- +(2,1.8) -- +(9.8,1.8) -- +(8,0) -- cycle; 
\node[hole] at (4,8.4) {};
\draw[journey] (4,8.4) -- (4,9.8);

\path[layer,filemanagerlayer] (0,9.6) -- +(2,1.8) -- +(9.8,1.8) -- +(8,0) -- cycle; 
\node[file,align=left] at (3.4,10.8) {source.c\phantom{MM}\\};
\node[file,align=left] at (6.2,10.8) {source.h\phantom{MM}\\};

\node[layerlabel] at (0.4,0.2) {CodeGen};
\node[layerlabel] at (0.4,1.8) {Sema};
\node[layerlabel] at (0.4,3.4) {Parser};
\node[layerlabel] at (0.4,5) {Preprocessor};
\node[layerlabel] at (0.4,6.6) {Lexer};
\node[layerlabel] at (0.4,8.2) {SourceManager};
\node[layerlabel] at (0.4,9.8) {FileManager};

\node[ast] at (-1.2,1.2) {AST};

%

\node[data] at (6,7.6) {Characters};
\node[data] at (6.2,7) {SourceLocation};
\node[data] at (5.8,5.8) {Tokens};
\node[data] at (5.8,4.2) {Tokens};
\node[data] at (6,9) {MemoryBuffer};
\end{tikzpicture}
\end{resizepar}
\caption{Clang's internal component layers}
\label{fig:frontend}
\end{figure}
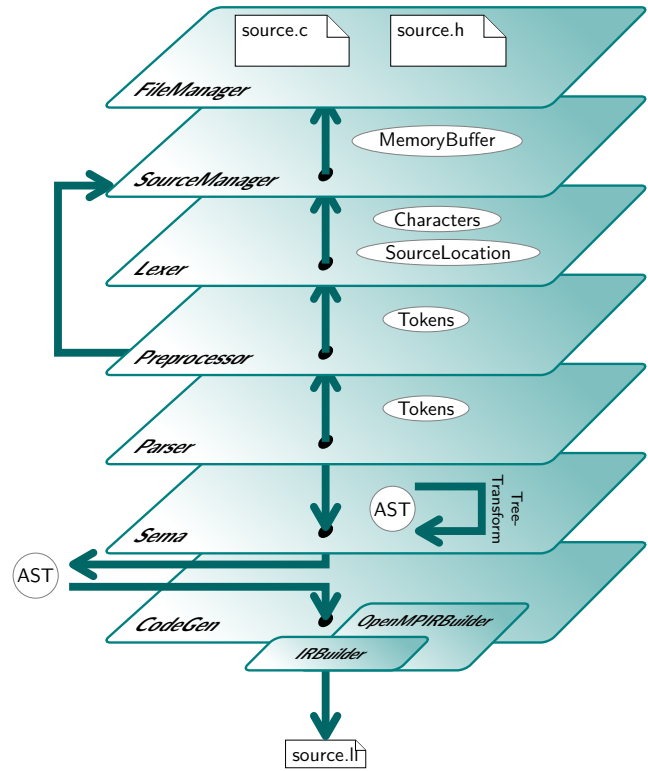

A compiler front-end is responsible for parsing source code, determine its meaning (semantics), and translate it into an intermediate representation (IR) designed to easy analysis an transformation that is (mostly) unspecific in regards to input programming language and target instruction set architecture.

Within the LLVM compiler infrastructure project~\cite{lattner02-llvm}, the front-end for C, C++ and Objective-C is Clang~\cite{clang}.
Clang 3.8 also added an implementation of OpenMP~\cite{llvmopenmp} in using an ``early outlining'' approach~\cite{bataev2014-openmp}.
That is, all OpenMP semantics are lowered in the front-end and the generated IR does not contain OpenMP-specific constructs, but calls to an OpenMP runtime.

\subsection{OpenMP Loop Transformation Directives}

OpenMP~5.1~\cite{openmp51} added loop nest transformations to the OpenMP language.
Before this change, OpenMP directives could only apply to statements that a programmer has written explicitly in the source code.
In the new OpenMP version, a loop transformation directive applied to a loop stands in for another loop as determined by the directive's definition.

In the example below, we first apply loop unrolling to the literal for-loop.
This results in another, unrolled, loop onto which another directive can be applied to; for instance, a \cinline{parallel for} directive:
\begin{minted}{c}
#pragma omp parallel for
#pragma omp unroll partial(2)
for (int i = 0; i < N; i+=1)
  body(i);
\end{minted}
The code above is semantically equivalent to the following version where the loop is unrolled manually by the programmer.
\begin{minted}{c}
#pragma omp parallel for
for (int i = 0; i < N; i+=2) {
  body(i);
  if (i+1 < N) body(i+1);
}
\end{minted}
As a result, transformations are applied in reverse order as they appear in the source code.
This is consistent with any other pragma that appear before the item they apply to.
With the addition of loop transformations, this can be either a \emph{literal loop} (by analogy with literal expression constants) that appears in the source code, or or a loop that is the result of a transformation, which we refer to as a \emph{generated loop}.

Such directives enable the separation of the semantics of algorithms and its performance-optimization~\cite{iwomp19-designanduse}.
For one, it improves the maintainability of the code: The directive clearly conveys the intend of the directives, compared to where the unrolling is intermingled with algorithm itself.
Using unrolling as an example, the \cinline{body} has to be duplicated multiply times.
If the unroll factor was to be changed, multiple expressions have to stay consistent with each other, including the \cinline{body} copies themselves, with any accidental inconsistency leading to potentially wrong results.
Hence, dedicated loop transformations make it easier to experiment with different optimization to find the best-performing on a particular hardware.
Moreover, different optimizations can be chosen for different hardware by either using the preprocessor, or the OpenMP \cinline{metadirective}, while using the same source code for the algorithm itself.

The implementation challenge is that before OpenMP~5.1 no directive was freely composable with other directives in arbitrary order and multiplicity.
There were only combined and composite directives with all valid combinations enumerated explicitly in the specification.
OpenMP~5.1 introduced two loop transformation directives: \cinline{tile} and \cinline{unroll}.
Tiling applies to multiple loops nested inside each other and generates twice as many loops.

Unrolling has a full, partial, and heuristic mode.
If fully unrolled, there is no generated loop that can be associated with another directive.
Partial unrolling can be understood as first tiling the loop by an unroll-factor, then fully unrolling the inner loop.
In heuristic mode, the compiler decides what to do: Full unroll, partial unroll with a chosen unroll factor, or not unroll at all.

\begin{figure}
\begin{minted}{c}
int i = 0;
for (; i+3 < N; i+=4)  { // unrolled
  body(i);
  body(i+1);
  body(i+2);
  body(i+3);
}
for (; i < N; i+=1)     // remainder
  body(i);
\end{minted}
\caption{Partial unrolling with remainder loop}\label{lst:remainder}
\end{figure}
A typical implementation of unrolling avoids the conditional within the loop and instead peels the last iteration into a remainder loop, as shown in \cref{lst:remainder}.
Implementations are allowed to apply this as an optimization as ling and the code's semantics are preserved.


\subsection{The Clang Abstract Syntax Tree}

An Abstract Syntax Tree (AST) is the structural in-memory representation of a program's source code.
Clang's AST mixes syntactic-only (such as parenthesis) and semantic-only (such as implicit conversions) nodes into the same tree structure.
With a few exceptions it is immutable, meaning that a subtree cannot be modified after it has been created.

\begin{figure}
\begin{subfigure}{\linewidth}
\begin{minted}{c}
#pragma omp parallel for schedule(static)
for (int i = 7; i < 17; i += 3)
  body(i);
\end{minted}
\caption{}\label{lst:astdump_a}
\end{subfigure}\newline
\begin{subfigure}{\linewidth}
\begin{scalepar}{0.69}
\begin{minted}{text}
OMPParallelForDirective
|-OMPScheduleClause
| `-[...]
`-CapturedStmt
  `-CapturedDecl nothrow
    |-ForStmt
    | |-DeclStmt
    | | `-VarDecl 0x7fffc6750e68 used i 'int' cinit
    | |   `-IntegerLiteral 'int' 7
    | |-[...]
    | |-[... (Cond)]
    | |-[... (Incr)]
    | `-CallExpr 'void'
    |   `-[...]
    |-ImplicitParamDecl implicit .global_tid. 'const int *const __restrict'
    |-ImplicitParamDecl implicit .bound_tid. 'const int *const __restrict'
    |-ImplicitParamDecl implicit __context '(unnamed struct) *const __restrict'
    `-VarDecl 0x7fffc6750e68
\end{minted}
\end{scalepar}
\caption{}\label{lst:astdump_b}
\end{subfigure}
\caption{An OpenMP loop-associated construct~(\subref{lst:astdump_a}) and its AST~(\subref{lst:astdump_b}) as printed by \cinline{clang -Xclang -ast-dump}; brackets indicate omissions from the raw output}
\label{lst:astdump}
\end{figure}

\Cref{lst:astdump} shows an example of an AST for an OpenMP directive associated to a for-loop.
The root of this subtree represents the \textinline{parallel for} pragma itself.
The child nodes at the beginning are the directive's clauses and their arguments, if any.

The last child node is the code the directive is associated with.
It is wrapped inside a \cinline{CapturedStmt} which borrows from Clang's C++ lambda and Objective-C's block implementation.
The \cinline{CapturedDecl} node contains the `lambda function' definition, \cinline{CapturedStmt} represents the statement that declares it and the \cinline{OMPParallelForDirective} is responsible for calling it.
Re-purposing the lambda/block implementation makes it easier to outline the directive's associated code into another function which is necessary to call it from other threads.
Clang also keeps track of which variables are used inside the \cinline{CapturedStmt} to become parameters of the outlined function.
In \cref{lst:astdump} these are indicated by the \cinline{ImplicitParamDecl} nodes for passing the thread identifiers, a context structure wrapping the captured variables, and the loop iteration variable itself.

The loop itself is represented by the \cinline{ForStmt}, the same AST node as if the loop was not part of an OpenMP directive.
It's children are the components of a C/C++ for-loop (initialization, condition, and increment) and its body, here a call to another function.
The iteration variable \cinline{VarDecl} capture of the \cinline{CapturedStmt} is in fact only a reference to the declaration in the for-loops init-statement.
A C++11 Range-Based For-Loop~\cite{cpp11} would be represented by a \cinline{CXXForRangeStmt}.
For convenience in the analysis, its children also include some of the statements that the range for-loop is equivalent to (``de-sugared'', see \cref{lst:unsugar,lst:rangesugar}), which has slightly changed between C++11, C++17 and C++20.
Ideally, such changes are abstracted over such that analysis code does not have to handle each standard separately.

\begin{figure}
\begin{scalepar}{0.7}
\begin{tightcenter}
\begin{tikzpicture}
\graph[layered layout] {
  Stmt -> Expr -> dot0[as=\dots];
  Stmt -> {ForStmt, CXXForRangeStmt};
  OMPExecutableDirective -> { OMPParallelDirective, dots1[as=\dots]};
  Stmt -> OMPExecutableDirective -> OMPLoopDirective -> {OMPForDirective, OMPParallelForDirective, dots2[as=\dots]};
  Stmt -> CapturedStmt;
};
\end{tikzpicture}
\end{tightcenter}
\end{scalepar}
\caption{Excerpt of the AST node class hierarchy}
\label{fig:ompclass}
\end{figure}
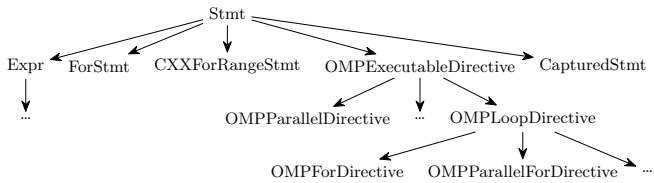

As shown in \cref{fig:ompclass}, \cinline{OMPParallelForDirective} is derived from \cinline{OMPLoopDirective}, a base class for all loop-associated directives.
The latter is derived from \cinline{OMPExecutableDirective} which is a base class for all OpenMP directives whose syntax allows them to be placed wherever a base language statement can appear.
Accordingly, it itself is derived from the \cinline{Stmt} class.
Declarations (\cinline{Decl}, such as \cinline{CapturedDecl}), types (\cinline{Type}) and clauses (\cref{fig:clauseclass}) are not related in the class hierarchy, i.e. there is no common base class for AST nodes.
Expressions on the other hand can be uses as a statement with its result being ignored, hence \cinline{Expr} is derived from \cinline{Stmt}.
For walking over all AST nodes, a visitor pattern separate for each of the type hierachies must be used (\cinline{StmtVisitorBase}, \cinline{DeclVisitor}, \cinline{TypeVisitor}, \cinline{OMPClauseVisitor}).

An \cinline{OMPExecutableDirective} may contain additional AST nodes that are not part of the AST node's \cinline{children()} enumeration\footnote{The inherited method \cinline{children()} returns a list of \cinline{Stmt}s, hence it cannot enumerate any \cinline{OMPClauses}. They are still printed in an AST dump using specialized functions fo nearly every AST node subclass.} and are not emitted in the AST dump such as in \cref{lst:astdump,lst:astdump_shadowast}.
We use the term \emph{Shadow AST} for such hidden children.
Presumably, this was done to not print excessive output, and/or avoid unintentionally referencing them by AST consumers and regression tests.

\cinline{OMPLoopDirective} has up to 30 shadow AST statements for representing a loop nest, plus 6 for each loop in the associated loop nest.
Like the \cinline{CXXForRangeStmt}'s de-sugared AST nodes these contain implicit code, but without these having been mandated by the OpenMP specification.
Examples of these nodes include: The expression to compute the number of iterations, whether an iteration is the last iteration, how to compute the next loop counter value, etc.
That is, a significant portion of the code generation already takes place when creating the AST.

\subsection{Clang Layer Architecture}

Clang's internal organization is sketched in \cref{fig:frontend}.
It follows a typical compiler structure consisting of tokenizer/Lexer, Preprocessor, Parser, semantic analyzer (Sema), and IR code generation (CodeGen).
General control flow is steered by the parser.
That is, when calling the parser's \cinline{ParseTopLevelDecl()}, it pulls the tokens to be consumed from the previous layers.
When the parser has decided what syntactic element it is, it is pushed to Sema to create an AST node for it.
Sema also performs the semantic analysis including creating implicit AST nodes.
The \cinline{TreeTransform} class creates copies of AST subtrees with some changes applied.
Its primary use is template instantiation: When instantiating or specializing a template, it creates a new AST subtree with substituted template parameters.

The result is a complete AST that must not be modified after this point.
It can used by tools such as source-to-source code generators, \emph{clang-tidy}, \emph{clang-query}, IDEs, \emph{include-what-you-use}, etc.
Since Clang is a compiler, its default action is to pass it to CodeGen, which produces functions and instructions for the mid-end to be optimized.
Although it is possible to emit diagnostics and errors in CodeGen, it is preferred to emit them in the semantic analyzer and the layers before, as tools not using CodeGen including Clang's own \cinline{-syntax-only} would otherwise not emit them.

When asked to generate IR for an OpenMP directive, the designated method decides how to emit IR instructions.
CodeGen's \cinline{EmitOMPParallelForDirective} method emits a new outlined function (the \cinline{#pragma omp parallel} part) with the calls to the OpenMP runtime that manages the threads, emits thread-number dependent conditionals (the \cinline{#pragma omp for} part), and emits the loops itself (common for all \cinline{OMPLoopDirective}-derived directives).
Since these parts are modular for OpenMP combined and composite directives, the actions are chained using callbacks, where each part can replace the body code generation function and call the previous callback (``callback-ception'').

The IR instructions themselves are emitted through \cinline{IRBuilder}, a class that offers many convenience functions to create any instruction, inserts them after the previously inserted instruction, attaches debug info, and offers a callback interface than can make modifications on just inserted instructions.
Additionally, it simplifies expressions (e.g. algebraic simplifications) on-the-fly which avoids creating instructions that would later be optimized away anyway.

A recent development is the introduction of the OpenMPIRBuilder~\cite{patch-openmpirbuilder} to extract out the base-language independent portion of the OpenMP lowering from the one that is specific to the Clang AST.
The goal is to share the implementation of the heavy lowering between Clang and the MLIR OpenMP Dialect~\cite{mlir-omp-dialect}, similar to how \cinline{IRBuilder} is used by many language front-ends and not just Clang.
The building blocks provided by OpenMPIBuilder can also be used by other parallel languages such as OpenACC~\cite{denny2018clacc}.
MLIR is also generated by Flang~\cite{scalpone2020-flang}, meaning this will enable a shared OpenMP code generation between C/C++ and Fortran.
As of writing of this paper, this refactoring is still in progress.
It can be enabled using the experimental flag \textinline{-fopenmp-enable-irbuilder}.
Eventually, the OpenMPIRBuilder will replace Clang's current CodeGen implementation for OpenMP.

As a result, we implemented two versions of loop transformation directives.
The first version (\cref{sct:shadowast}) is following the shadow AST approach which is compatible with the current approach.
The second version (\cref{sct:canonicalloop}) implements the base-language invariant parts in the OpenMPIRBuilder and moving as much of the code generation from the Sema to the CodeGen layer.
This gives the opportunity to share the implementation with Fortran and to refactor the current AST modeling.

\section{Shadow AST Representation}\label{sct:shadowast}

The idea behind this implementation is to apply the transformation on the loops in the AST, creating a new AST, similar to how \cinline{TreeTransform} works already.
This has the advantage that one can choose, depending on the operation, to either use the AST representing the parsed code, or the AST that represents the semantics.
When normally accessing or printing the AST, only the parsed/syntactical AST is returned while the transformed AST is a shadow AST.
Unlike a \cinline{CXXForRangeStmt}, the entire de-sugared statement is stored, not just some individual statements~\cite{patch-shadowast-unroll,patch-shadowast-tile}.

When another directive is applied to the loop transformation AST node, it calls \cinline{getTransformedStmt()} to get the semantically equivalent AST.
This has the advantage that the existing code of the directive for analyzing the loop nest does not need to be changed and applies to the transformed code as if it was a literal for-loop.

Some care needs to be taken for compiler diagnostics: The existing semantic analysis assumes that the AST nodes represent code from the source file, but it may accidentally refer to the internal shadow AST.
For instance, after tiling there is a loop variable for the inner and one for the outer loop.
If a diagnostic prints the variable name, the user will see a diagnostic such as
\begin{minted}{text}
note: read of non-const variable '.capture_expr.' is
      not allowed in a constant expression
\end{minted}
which is not useful to the programmer.
If the diagnostic only points to a \cinline{SourceLocation}, a representative source location for the associated literal loop can be used, even though the diagnostic applies to the transformed AST and may not make sense one the original AST.
A ``note'' diagnostic\footnote{Note diagnostics augment warning and error diagnostics with additional relevant source locations, such as ``template instantiation required here''.} for explaining the history of the location similar to template instantiation and macros expansion might be useful to improve the quality of the implementation.

\subsection{Abstract Syntax Tree Changes}

\begin{figure}
\begin{scalepar}{0.7}
\begin{tightcenter}
\begin{tikzpicture}
\graph[layered layout] {
  OMPExecutableDirective -> { OMPParallelDirective, dots1[as=\dots] };
  Stmt -> OMPExecutableDirective -> OMPLoopBasedDirective[text=red] -> OMPLoopDirective -> {OMPForDirective, OMPParallelForDirective, dots2[as=\dots]};
  OMPLoopBasedDirective -> {OMPUnrollDirective[text=darkgreen], OMPTileDirective[text=darkgreen]};
};
\end{tikzpicture}
\end{tightcenter}
\end{scalepar}
\caption{Stmt class hierarchy for loop transformations}
\label{fig:shadowastclass}
\end{figure}
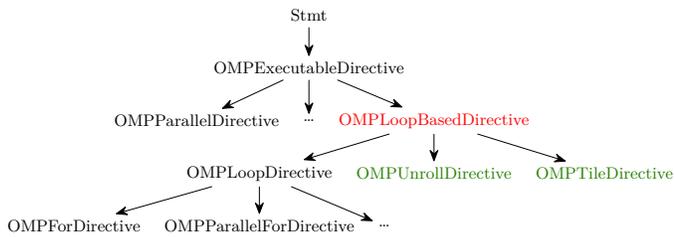

The changes to the AST's Stmt class hierarchy are shown in \cref{fig:shadowastclass}.
Most straightforwardly, two classes \cinline{OMPUnrollDirective} and \cinline{OMPTileDirective} have been added, representing an unroll pragma, respectively a tile pragma.
But they are derived from \cinline{OMPLoopBasedDirective}, a new class inserted between \cinline{OMPExecutableDirective} and \cinline{OMPLoopDirective}.
The motivation is that only the transformed AST is needed for loop transformations, but not the many other shadow AST nodes that \cinline{OMPLoopDirective} comes with.
For instance, the expression that computes the number of iterations is only needed during the construction of the transformed AST and will be part thereof.
But it is not needed separately by CodeGen or any other layer once it has been created.
Any directive applied to the transformed loop will (re-)analyze the transformed AST without needing access to intermediate steps.
As a drawback, the transformed AST must an OpenMP canonical loop nest itself or otherwise will be rejected by that analysis.

\begin{figure}
\begin{scalepar}{0.7}
\begin{tightcenter}
\begin{tikzpicture}
\graph[layered layout] {
  OMPClause -> {OMPFullClause[text=darkgreen],OMPPartialClause[text=darkgreen],OMPSizesClause[text=darkgreen],dots[as=\dots]};
};
\end{tikzpicture}
\end{tightcenter}
\end{scalepar}
\caption{OpenMP clause class hierarchy}
\label{fig:clauseclass}
\end{figure}
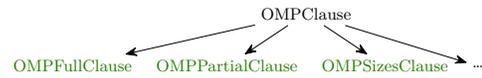

The extended class hierarchy for clauses is shown in \cref{fig:clauseclass}, but with not surprises here.

\begin{figure}
\begin{subfigure}{\linewidth}
\begin{minted}{c}
#pragma omp unroll full
#pragma omp unroll partial(2)
for (int i = 7; i < 17; i += 3)
  body(i);
\end{minted}
\caption{}\label{lst:astdump_shadowast_a}
\end{subfigure}\newline

\begin{subfigure}{\linewidth}
\begin{scalepar}{0.69}
\begin{minted}{text}
OMPUnrollDirective
|-OMPFullClause
`-OMPUnrollDirective
  |-OMPPartialClause
  | `-ConstantExpr 'int'
  |   |-value: Int 2
  |   `-IntegerLiteral 'int' 2
  `-ForStmt
    |-DeclStmt
    | `-VarDecl used i 'int' cinit
    |   `-IntegerLiteral 'int' 7
    |-[...]
    |-[... (Cond)]
    |-[... (Incr)]
    `-CallExpr 'void'
      `-[...]
\end{minted}
\end{scalepar}
\caption{}\label{lst:astdump_shadowast_b}
\end{subfigure}
\caption{Composition of two directives~(\subref{lst:astdump_shadowast_a}) and its AST~(\subref{lst:astdump_shadowast_b})}
\label{lst:astdump_shadowast}
\end{figure}

A transformed loop can itself again become subject of a loop transformation as demonstrated in \cref{lst:astdump_shadowast}.
Its loop is first partially unrolled, then fully unrolled, which is effectively equivalent to just being unrolled completely.

Unlike directives derived from \cinline{OMPLoopDirective}, the loop body code is not wrapped inside a \cinline{CapturedStmt}.
There is no need for it because it will never be outlined unless nested inside another region that is outlined.
For loop transformations themselves it is imperative to not wrap the code in a \cinline{CapturedStmt} because local variables are changed to refer the \cinline{CapturedStmt}'s implicit parameters which the transformed AST would have to either replicate or changed back.

\begin{figure}
\begin{scalepar}{0.69}
\begin{minted}{text}
ForStmt
|-[int unrolled.iv.i = 3]
|-[...]
|-[unrolled.iv.i < (17-(7-3+1))/3]
|-[unrolled.iv.i += 2]
`-AttributedStmt
  |-LoopHintAttr Implicit loop UnrollCount Numeric
  | `-IntegerLiteral 'int' 2
  `-ForStmt
    |-[int unroll_inner.iv.i = unrolled.iv.i]
    |-<<<NULL>>>
    |-unroll_inner.iv.i < unrolled.iv.i + 2 && unroll_inner.iv.i < 4]
    |-[++unroll_inner.iv.i]
    `-[... (body)]
\end{minted}
\end{scalepar}
\caption{Transformed AST of the unroll directive in \cref{lst:astdump_shadowast}}
\label{lst:transformedast}
\end{figure}

\Cref{lst:transformedast} shows part of the shadow AST for the inner \cinline{OMPUnrollDirective} in \cref{lst:astdump_shadowast}.
The loop has been strip-mined using a tile size of 2.
Instead of cloning the body statement according to the unroll factor, the inner loop is kept and annotated with a \cinline{LoopHintAttr} attribute, the same as used by
\begin{minted}[escapeinside=``]{c}
#pragma clang loop unroll_count(2) `\textcolor{black}{.}`
\end{minted}
Upon encountering this attribute, the code generator will attach \cinline{llvm.loop.unroll.count} metadata to the node which is interpreted by the \cinline{LoopUnroll} pass in the mid-end to eventually unroll the loop.
No duplication takes place until that point.
\cinline{LoopUnroll} will also handle the case when the iteration count is not a multiple of the unroll factor.

\subsection{Code Generation Changes}

A transformed AST is only necessary if the replacement is potentially associated with another directive, which according to OpenMP rules is only possible if the \cinline{partial} clause is present.
Hence, the outer directive of \cref{lst:astdump_shadowast} does not have a shadow AST.
Instead, CodeGen emits its IR directly.
Otherwise, the consuming directive has analyzed the (new) loop bounds and becomes responsible for its code generation.

If encountering a non-associated tile construct, CodeGen will simply emit the transformed AST in its place.
For the unroll directive, it is more efficient to defer unrolling to the \cinline{LoopUnroll} pass by attaching \cinline{llvm.loop.unroll.*} metadata to the loop without even tiling the loop beforehand.
This has the additional advantage that, if the compiler is allowed to choose the unroll factor itself, the \cinline{LoopUnroll} pass can apply profitability heuristics to determine an appropriate factor.

If the unrolled loop is consumed by another directive, the unroll factor must be chosen without \cinline{LoopUnroll}'s heuristic because it is already used in shadow AST.
The unroll factor determines the number of iterations of the unrolled loop and can become observable when associated by another directive, such as the \cinline{taskloop} creating as many task as there are iterations.
The current implementation~\cite{patch-shadowast-unroll} uses the unroll factor of two in this case.
Future improvements may implement a better heuristic.

\section{Canonical Loop Representation}\label{sct:canonicalloop}

The idea behind this implementation is to move as much code generation logic as possible from the Sema layer into the CodeGen layer, specifically into OpenMPIRBuilder such that it can be shared between Clang and Flang.
Unfortunately, significant parts of the code generation result is stored in the shadow AST of \cinline{OMPLoopDirective}.
However, not all of the loop analysis can be moved into the CodeGen layer: We still want to diagnose malformed loops in Sema, and --- even more importantly, some constructs are inherently base-language dependent.
In C++ a loop over iterators or a range-based for-loop requires overload resolution and potentially template instantiation for expressions that do not literally appear in the source code.
For instance, the expression $ub-lb$ to compute the distance between loop start and loop end, where $lb$ and $ub$ are iterator classes, requires resolving the correct overload of the subtraction operator.

Instead we are abstracting the loop iterations variable~\cite{patch-clang-createWorkshareLoop}.
That is, internally (as already in the OpenMP runtime) the \emph{logical iteration counter} is always a normalized unsigned integer starting at 0 and incrementing by one at each iterations.
Its value therefore corresponds to the logical iteration number used in the OpenMP specification.

We call the variable that a literal for-loop uses for keeping track of the iterations the \emph{loop iteration variable}.
A generated loop does not have a loop iteration variable, whether the IR emitted by CodeGen or starts at 0 or any other number is irrelevant and subject to mid-end optimizations anyway.

\begin{figure}
\begin{subfigure}{\linewidth}
\begin{minted}{cpp}
for (double &Val : Container)
  body(Val);
\end{minted}
\caption{}\label{lst:unsugar}
\end{subfigure}\newline

\begin{subfigure}{\linewidth}
\begin{minted}{cpp}
auto &&__range = Container;
auto __begin = std::begin(__range);
auto __end = std::end(__range);
for (; __begin != __end; ++__begin) {
  double &Val = *__begin;
  body(Val);
}
\end{minted}
\caption{}\label{lst:rangesugar}
\end{subfigure}\newline

\begin{subfigure}{\linewidth}
\begin{minted}[linenos]{cpp}
auto &&__range = Container;
auto __begin = std::begin(__range);
auto __end = std::end(__range);
size_t Distance = std::distance(__begin,__end);
for (int __i = 0; __i < Distance; ++__i) {
  double &Val = *(__begin + __i);
  body(Val);
}
\end{minted}
\caption{}
\label{lst:rangeloop_desugared}
\end{subfigure}
\caption{Three implementations of loop at various stages of de-sugaring; \texttt{Val} is the \emph{loop user variable}, \texttt{\_\_begin} is the \emph{loop iteration variable}, and \texttt{\_\_i} is the \emph{logical iteration counter}}
\label{lst:rangeloop}
\end{figure}

Third, the \emph{loop user variable} is the user-accessible variable that the loop body code depends on and may have a different value in each iteration.
For a literal for-loop it is identical to the loop iteration variable, but in range-based for-loop the iterator itself is inaccessible by user code and only the dereferenced iterator can be used within body code.
For an illustration, see \cref{lst:rangeloop}.
Note that \cref{lst:rangeloop_desugared} is only semantically equivalent to \cref{lst:rangesugar,lst:unsugar} if the loop fulfills OpenMP's canonical loop constraints.

We identified the following minimal set of meta-information that need to be resolved at the Sema-layer:
\begin{enumerate}
\item Distance function: An expression evaluable before entering the loop for the loop trip count (Line~4 in \cref{lst:rangeloop_desugared}).
\item User value function: An expression to convert a logical iteration number into a value for the loop user variable (Line~6).
\item User variable reference: The user variable that needs to be updated before each iteration.
\end{enumerate}
This is reduced from the 36 shadow AST nodes required by \cinline{OMPLoopDirective}.

\subsection{Abstract Syntax Tree Changes}

Only one additional class derived from \cinline{Stmt} is introduced: \cinline{OMPCanonicalLoop}.
The other classes from \cref{fig:shadowastclass} introduced for loop transformations are reused.
\cinline{OMPCanonicalLoop} acts like an implicit AST node similar to an implicit cast.
It is inserted as the parent of a literal for-loop whenever it needs to be ``converted'' into an OpenMP canonical loop as part of a loop-associated directive and can be losslessly removed again if the wrapped loop needs to be re-analyzed.

\begin{figure}
\begin{scalepar}{0.69}
\begin{minted}{text}
OMPUnrollDirective
`-OMPCanonicalLoop
  |-ForStmt
  | |-[... (init)]
  | |-<<<NULL>>>
  | |-[... (cond)]
  | |-[... (incr)]
  | `-CallExpr 'void'
  |   `-[...]
  |-CapturedStmt
  | `-[... (distance)]
  |-CapturedStmt
  | `-[... (loop value)]
  `-DeclRefExpr 'int' lvalue Var 'i' 'int'
\end{minted}
\end{scalepar}
\caption{Unroll directive using \texttt{OMPCanonicalLoop}}
\label{lst:ompcanonicalloop}
\end{figure}

\Cref{lst:ompcanonicalloop} shows an example of an AST using \cinline{OMPCanonicalLoop}.
The first child is the loop (\cinline{ForStmt} or \cinline{CXXForRangeStmt}) it is wrapping.
The distance and loop user value functions are lambdas represented by \cinline{CaptureStmt} nodes.
Wrapping these expressions in lambdas is necessary to allow CodeGen to call them with any argument.
An \cinline{Expr} tree references concrete variables that cannot be changed after Sema.

The distance function has the following signature:
\begin{minted}{cpp}
[&](size_t &Result) {
  Result = __end - __begin;
}
\end{minted}
This sets the \cinline{Result} argument to the loop's trip count.
Here we are using \cinline{size_t} for the logical iteration count type, but it actually depends on the precision of the type of subtract expression, e.g. \cinline{ptrdiff_t} for pointers and most iterators.

\cinline{__end} and \cinline{__begin} are implicitly captured by reference.
These are not necessarily the variables introduced in \cref{lst:rangesugar}, but more generally the loop iteration variable after the for-loop's init statement has been executed: In other words, the loop iteration variable's start value.
Similarly, \cinline{__end} is the loop's upper bound.

Additional complexity may be necessary, such as evaluating to 0 if \cinline{__begin} is larger than \cinline{__end}, unless iterating in reverse.
Also special care must be taken to allow the maximum number of iterations.
For instance,
\begin{minted}{cpp}
for (int32_t i = INT32_MIN; i < INT32_MAX; ++i)
\end{minted}
has \cinline{0xfffffffe} iterations that do not fit into a 32-bit signed integer and therefore a reason why we always an unsigned logical iteration counter.
The number of iterations cannot be negative, but the trip count will never be equal to or exceed the range of an unsigned integer of the same bitwidth.
An iteration count of \cinline{0xffffffff} with 32 bit integer iteration variables is theoretically expressible in Fortran, but OpenMPIRBuilder does not support it.

The loop user value function's signature is the following:
\begin{minted}{cpp}
[&,__begin](auto &Result, size_t __i) {
  Result = __begin + __i;
}
\end{minted}
Again, the result is stored in a variable passed by-reference.
The result cannot be returned using the lambda's return value because that would be an r-value of a user-defined type.
It may trigger language-dependent overloads to copy/move its value into a memory location which can only be done in Sema.
By passing by reference (even in C), the memory referenced by \cinline{Result} will just have the expected bit pattern after the call returns, including having called the destructor for the previous value if necessary.

Captures take place before the loop itself, but it is evaluated inside the loop.
\cinline{__begin} is captured by-value so at any time it will contain the start value of the loop iteration variable even though it will be modified inside the loop.

When eventually Clang switches completely to OpenMPIRBuilder and removes the \cinline{OMPLoopDirective}-based implementation, all loop-associated directives can be changed to derive from \cinline{OMPLoopBasedDirective} instead and no transformed AST node need to be generated anymore.
While the \cinline{OMPUnrollDirective} does not wrap its associated code into a \cinline{CapturedStmt}, other directives such as \cinline{OMPParallelForDirective} still may.
They may also become unnecessary with further adaption of OpenMPIRBuilder which outlines on the IR-level instead of depending on the front-end to outline itself.

As a downside, without the transformed shadow AST, the semantic analyzer will need its own logic to verify that a loop nest after transformations is sufficiently deep to apply loop-associated directives.
For the moment we relay on the existing diagnostic that comes with the shadow AST implementation.

\subsection{Code Generation Changes}

\begin{figure}
\includegraphics[width=\linewidth]{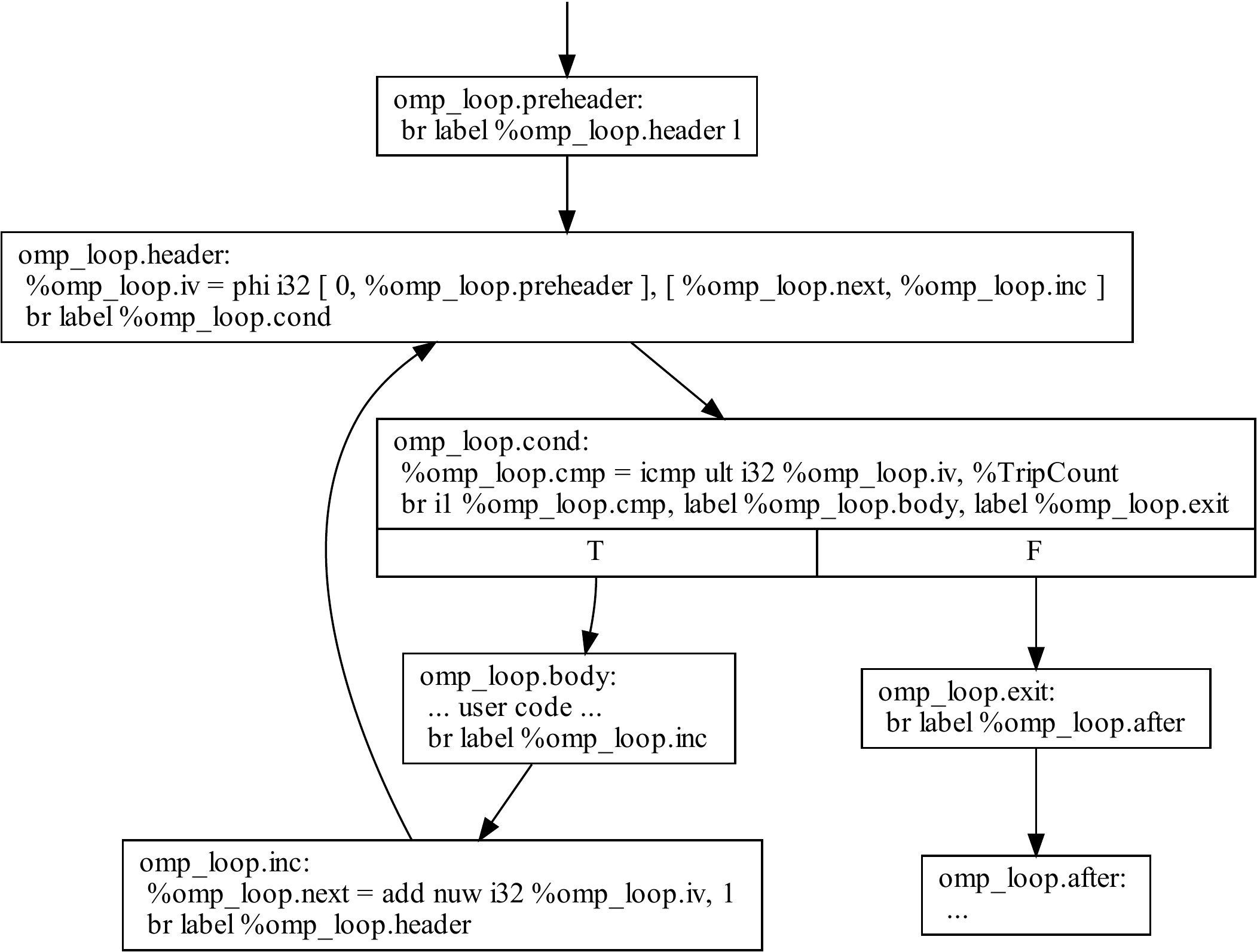}
\caption{Loop skeleton generated by \texttt{createCanonicalLoop}}
\label{fig:loopskeleton}
\end{figure}

When CodeGen has to emit an \cinline{OMPCanonicalLoop}, instead of using Clang's functions to emit a \cinline{ForStmt} or \cinline{CXXForRangeStmt}, it calls OpenMPIRBuilder's \cinline{createCanonicalLoop} function~\cite{patch-ompbuilder-createCanonicalLoop} which creates a loop skeleton in LLVM-IR (shown in \cref{fig:loopskeleton}).
It takes the loop's trip count as argument which CodeGen can get by calling the distance function, and returns a \cinline{CanonicalLoopInfo} object which represents the loop and its current state in the IR.
Among other information, it stores which \cinline{llvm::Value} represents the logical iteration counter and the location of the loop's body\footnote{For re-entry into callback-ception, \cinline{createCanonicalLoop} also takes a function argument.
\cinline{createCanonicalLoop} calls it with those two as arguments before it itself returns.}.
This is where CodeGen emits the \cinline{ForStmt}/\cinline{CXXForRangeStmt}'s body code.
Before that, it will call the loop user value function using the logical iteration counter to fill the loop user variable with content.

The \cinline{OMPCanonicalLoop} can also be used as a handle to pass to other functions such as \cinline{createWorkshareLoop}~\cite{patch-ompbuilder-createStaticWorkshareLoop} which implements the worksharing-loop construct, \cinline{tileLoops}~\cite{patch-ompbuilder-tileLoops} which implements the tile loop transformations, or \cinline{collapseLoops}~\cite{patch-ompbuilder-collapseLoops}.

In the case of loop transformations, the methods again return (one or more) \cinline{CanonicalLoopInfo}s that can in turn again be used as handles.
The function may either modify and return the input canonical loops, or abandon the old handles and create new loops using the skeleton.
In either case, returned loops must adhere to the loop skeleton invariants which include:
\begin{itemize}
\item Explicit basic blocks for preheader, header, condition check, body entry, latch, exit and after.
\item Identifiable logical iteration variable/induction variable.
\item Identifiable loop trip count, without requiring analysis by ScalarEvolution.
\end{itemize}

\section{Conclusion}

The shadow AST approach has already been implemented~\cite{patch-shadowast-tile,patch-shadowast-unroll} and works (modulo bugs) in the top-of-tree of Clang's development repository which eventually will become Clang 13.0.0.
The OpenMPIRBuilder implementation for handling loops in general~\cite{patch-ompbuilder-createCanonicalLoop} including its use by Clang~\cite{patch-clang-createWorkshareLoop} is still in active development and will need some time before becoming production-ready.
As of this writing, Clang is missing implementations for any loop-associated directive other than workshare-loop, any clauses other than the \cinline{schedule} clause, loop nests with more than one loop, use within templates, cancellation, exceptions, etc.
However, its advantages are a shared implementation with other front-ends such as Flang, and a simplification of the AST representation including the removal of ``hidden'' shadow AST subtrees and wrapping associated statement into \cinline{CapturedStmt}s.

As far as we know, no other compiler has yet implemented OpenMP loop transformations.
Since multiple vendor compilers are derived from Clang, it is expected that these will inherit the implementations described here.

OpenMP~6.0 is expected to introduce additional loop transformations and mechanisms to apply them to not just the outermost generated loop.
For example, after tiling a loop, it is possible to apply worksharing to the outer loop and \cinline{simd} to the inner loop.
Currently, only the former is possible.
Some directives may also be redefined as loop transformations.
For instance, further loop directives may be applicable to \cinline{simd}-generated loops.
The additional loop transformation will likely include loop fusion and fission
that handle sequences of loops in addition to loop nest.
The additional abstractions provided by the \cinline{OMPCanonicalLoop} AST node and the OpenMPIRBuilder build the foundation for implementing these extensions in Clang.

\begin{acks}
This research was supported by the Exascale Computing Project (17-SC-20-SC), a collaborative effort of the U.S. Department of Energy Office of Science and the National Nuclear Security Administration, in particular its subproject SOLLVE.

This research used resources of the Argonne Leadership Computing Facility, which is a DOE Office of Science User Facility supported under Contract DE-AC02-06CH11357.

\end{acks}

\bibliographystyle{ACM-Reference-Format}
\bibliography{bibliography}

\appendix

\end{document}